\documentclass[aps,pre,reprint,preprintnumbers]{revtex4-2} 
\usepackage{hyperref}
\usepackage{amsmath,latexsym}
\usepackage{graphicx}
\usepackage{siunitx}
\usepackage{textcomp}
\usepackage{wasysym}
\usepackage{amssymb}
\usepackage{MnSymbol}
\usepackage{oplotsymbl}
\usepackage{physics}
\usepackage{array}
\usepackage{tikz}

\DeclareSIUnit{\molar}{M}
\begin{document}
\title{Two Modes of Cluster Dynamics Govern the Viscoelasticity of Colloidal Gels}
\date{\today}
\author{Jae Hyung Cho}
\email{jaehcho@mit.edu}
\affiliation{Department of Mechanical Engineering, Massachusetts Institute of Technology, Cambridge, Massachusetts 02139, USA}
\author{Irmgard Bischofberger}
\email{irmgard@mit.edu}
\affiliation{Department of Mechanical Engineering, Massachusetts Institute of Technology, Cambridge, Massachusetts 02139, USA}

\begin{abstract}
Colloidal gels formed by strongly attractive particles at low particle volume fractions are composed of space-spanning networks of uniformly sized clusters. We study the thermal fluctuations of the clusters using differential dynamic microscopy by decomposing them into two modes of dynamics, and link them to the macroscopic viscoelasticity via rheometry. The first mode, dominant at early times, represents the localized, elastic fluctuations of individual clusters. The second mode, pronounced at late times, reflects the collective, viscoelastic dynamics facilitated by the connectivity of the clusters. By mixing two types of particles of distinct attraction strengths in different proportions, we control the transition time at which the collective mode starts to dominate, and hence tune the frequency dependence of the linear viscoelastic moduli of the binary gels.
\end{abstract}

\maketitle

\footnotetext[1]{See Supplemental Material for sample videos used for DDM.}
\footnotetext[2]{Estimated attraction strengths of the two types of particles based on their $\phi$ dependence of the storage modulus~$G'$ are $30k_{B}T$ and $3.5k_{B}T$ at $T=30\si{\celsius}$, where $k_{B}$ is the Boltzmann constant, under the assumption of the van der Waals potential \cite{Calzolari2017}. However, we expect significant noncentral forces between bonded particles due to the roughness of the shells \cite{Dingenouts1998}, which limits the accuracy of these estimates.}

\section{Introduction}
Whether a soft material behaves more like an elastic solid or a viscous fluid depends on the relative timescales of its microscopic dynamics and those of the applied strains or stresses \cite{Bird1987a}. Tuning the viscoelasticity of a complex fluid, therefore, requires understanding of the dynamic interactions among its constituents. Indeed, when coupled to their surroundings, tracer particles in systems like crosslinked polymer gels \cite{Gittes1997,Sprakel2007,Godec2014,Kumar2019}, dense emulsions \cite{Mason1995}, and entangled polymer solutions \cite{Mason1995,Xu1998,Sprakel2007,Guo2012} are known to exhibit cooperative thermal fluctuations, which reflect the macroscopic viscoelasticity. \par

Composed of space-spanning networks of uniformly sized aggregate units, or clusters, dilute colloidal gels should also display viscoelastic interactions among the clusters in their microscopic dynamics. Current models, however, typically include only two steps, neither of which is linked to the viscoelasticity. On short timescales, localized elastic fluctuations occur, as shown by the decrease in the temporal correlation function to a finite plateau \cite{Krall1997,Krall1998,Romer2000,Segre2001,Manley2005,Liu2013,Romer2014,Calzolari2017,Cho2020}. On long timescales, cooperative structural rearrangements cause a full decay of the correlation function, allowing the kinetically arrested systems to age \cite{Cipelletti2000,Duri2006,Trappe2007,Guo2011,Gao2015,Bouzid2017,Chaudhuri2017}. Although these fast and slow relaxation processes adequately describe the elasticity and the aging behavior of colloidal gels, respectively, they cannot account for the abundant experimental evidence that hints at their viscoelasticity, including the absence of well-defined intermediate plateaus in the correlation functions \cite{Krall1998,Romer2000,Segre2001,Liu2013,Calzolari2017,Cho2020} and the frequency dependence in the viscoelastic spectra over multiple decades \cite{deRooij1994,Eberle2012,Helgeson2014,Calzolari2017,Colombo2017,Aime2018b,Szakasits2019}. \par

In this work, we use differential dynamic microscopy (DDM) \cite{Cerbino2008,Giavazzi2009} and macroscopic rheometry to show that the cluster dynamics of dilute colloidal gels reflects the macroscopic linear viscoelasticity through the superposition of a localized and a collective mode. At early times, the mean squared displacement (MSD) of a cluster, unaffected by the slower fluctuations of its surrounding network, approaches a finite plateau, which represents the elasticity of the gel. At late times, the collective motion of the clusters exhibits subdiffusion with an exponent close to $0.6$ that gives rise to the viscoelasticity. These quasiequilibrium collective fluctuations precede the nonequilibrium aging dynamics that also displays long-ranged correlated motions \cite{Cipelletti2000,Bouzid2017,Chaudhuri2017}. By mixing two types of particles of different attraction strengths in varying proportions, but at a constant total particle volume fraction $\phi$, we control the mean attraction strength, thus changing the MSD during the early elastic response. The late viscoelastic response, however, is largely independent of the mixing proportion, which allows us to adjust the time at which the subdiffusion starts to dominate, and hence to tune the viscoelastic spectra of the binary gels. We suggest that this collective dynamics is mediated by the steric hindrance among the mutually constrained clusters, which induces the power-law exponent close to $0.6$ independent of $\phi$, both in the MSD and the linear viscoelastic moduli. \par 

\section{Methods}
\subsection{Particle synthesis and characterization}
We utilize polystyrene-poly(N-isopropylacrylamide) (PS-PNIPAM) core-shell particles of two different PNIPAM shell thicknesses, synthesized via emulsion polymerization \cite{Dingenouts1998,Calzolari2017,Cho2020}. For the synthesis of the particles with thinner shells, we follow the protocol described in Ref. \cite{Calzolari2017}, which is slightly modified from that of Ref. \cite{Dingenouts1998}. In a $1\;\si{\liter}$ flask equipped with a stirrer, a reflex condenser, and a gas inlet, $25.02\;\si{\gram}$ of N-isopropylacrylamide (NIPAM, Acros Organics) and $0.2008\;\si{\gram}$ of the stabilizer sodium dodecyl sulfate (SDS, Sigma-Aldrich) are dissolved in $525.14\;\si{\gram}$ of DI water. After the solution is bubbled with nitrogen for $30\;\si{\min}$, $142.75\;\si{\gram}$ of styrene (Sigma-Aldrich) is added, and the mixture is heated to $80\si{\celsius}$ in nitrogen atmosphere. Then $0.3521\;\si{\gram}$ of the initiator potassium persulfate (KPS, Acros Organics) dissolved in $15.00\;\si{\gram}$ of DI water is added to the mixture. After $6\;\si{\hour}$, the dispersion is cooled to room temperature and cleaned through repeated centrifugation and supernatant exchange. For the synthesis of the particles with thicker shells, we follow an additional step of the seeded emulsion polymerization in Ref. \cite{Dingenouts1998} that increases the thickness of the PNIPAM shell, with slight modification of the ratio of the materials. For each $100\;\si{\gram}$ of the particles obtained from the first part, $12.58\;\si{\gram}$ of NIPAM and $0.8994\;\si{\gram}$ of the crosslinker N,N'-methylenebis(acrylamide) (BIS, Sigma-Aldrich) are added, and the mixture is heated to $80\si{\celsius}$. After the addition of $0.1264\;\si{\gram}$ of KPS dissolved in $9.43\;\si{\gram}$ of DI water, the mixture is stirred for $4\;\si{\hour}$. The suspension is then cooled to room temperature, and cleaned by dialysis against DI water for approximately four weeks. \par

To prevent sedimentation, we density-match all samples using a H\textsubscript{2}O/D\textsubscript{2}O mixture of 52/48 v/v. To minimize the effect of electrostatic interactions, we add $0.5\;\si{\molar}$ of sodium thiocyanate (NaSCN) to screen the charges of the particles \cite{Cho2020}. \par  

At temperatures $T$ lower than the gelation temperatures $T_{g}$, the thermosensitive shells induce sufficiently long-ranged steric repulsions stabilizing the particles. Both types of shells, however, are negligibly thin compared to the PS cores, such that when the shells shrink with increasing $T$, the particles aggregate by van der Waals attraction. We estimate the gelation temperatures $T_{g}$ by measuring the temperature at which the storage modulus $G'$ becomes larger than the loss modulus $G''$ at a frequency $\omega=6.28\;\si{\radian\per\s}$ during a temperature ramp experiment at a ramp rate of $0.2\;\si{\celsius\per\min}$, sufficiently slow to ensure uniform sample temperature \cite{Calzolari2017}. For the particles with the thinner and the thicker shells, $T_{g}=27.3\si{\celsius}$ and $25.5\si{\celsius}$, respectively. Since both types of particles are made of dense PS cores covered in much thinner PNIPAM shells \cite{Dingenouts1998,Calzolari2017}, we measure the volume fraction of our samples from the changes in their mass after drying in an oven, and assuming that the particle density is equal to the density of polystyrene $\rho=1.05\;\si{\g\per\cubic\centi\meter}$.\par

At any $T>27.3\si{\celsius}$, the particles with the thinner shells attract one another more strongly than those with the thicker shells because of the shorter-ranged repulsions, and thus we term them the strong particles, and the other set the weak particles \cite{Note2}. For their mixtures at any $\phi$, we denote the volume proportion of the strong particles by $\chi\in[0, 1]$. The hydrodynamic radii $a$ measured via dynamic light scattering (BI-200SM, Brookhaven Instruments) at $T=30\si{\celsius}$, at which all our experiments are conducted, are $90.3\pm1.6\;\si{\nano\meter}$ and $116.3\pm1.8\;\si{\nano\meter}$ for the strong and the weak particles, respectively. This small difference in $a$ is insignificant compared to the changes in the dynamic and the rheological parameters reported here. Since all the measurements are taken at $T=30\si{\celsius}$, the temperature dependence of the attraction strength between the particles does not play a role in our findings. \par

\subsection{Rheometry and differential dynamic microscopy}
To measure the linear viscoelasticity of the gels both macroscopically and microscopically, we employ oscillatory rheometry and DDM. For gels at $\phi=5.0\%$, the storage modulus $G'$ and the loss modulus $G''$ at different frequencies $\omega$ are obtained with a cone-plate geometry (diameter: $40\;\si{\milli\meter}$) on a stress-controlled rheometer (AR-G2, TA Instruments). For gels at lower $\phi=0.5 - 2.5\%$, we measure the density fluctuations of the gels in terms of the normalized intermediate scattering function $f(t,q)$, where $t$ denotes the delay time and $q$ the wave vector, as described in Appendix~\ref{ddm_detail}. We load the samples in glass capillary tubes (Vitrocom) of thickness $100\;\si{\micro\meter}$, and use an sCMOS camera (Prime Mono, $2048\times2048$ pixels, Photometrics) mounted on an inverted microscope (Eclipse TE2000-U, Nikon) with a water immersion objective of magnification $M=60\times$ and numerical aperture $\mathrm{NA}=1.20$. To probe the dynamics over a broad range of timescales, we use three stacks of 1000 frames acquired at different frame rates ranging from $0.5\;\mathrm{fps}$ to $100\;\mathrm{fps}$. For both rheometry and DDM, we thoroughly mix the strong and the weak particles in stable suspensions, and then initiate the gelation at time $t_w=0\;\si{\s}$ by rapidly increasing the temperature from $20$ to $30\si{\celsius}$. We start the measurements after $t_w=1800\;\si{\s}$ for rheometry and $t_w=5400\;\si{\s}$ for DDM, such that aging effects are negligible during the data acquisition. \par

\section{Two modes of cluster dynamics and their link to viscoelasticity}

\begin{figure*}[t]
\setlength{\abovecaptionskip}{10pt}
\hspace*{-0.06cm}\includegraphics[scale=0.31]{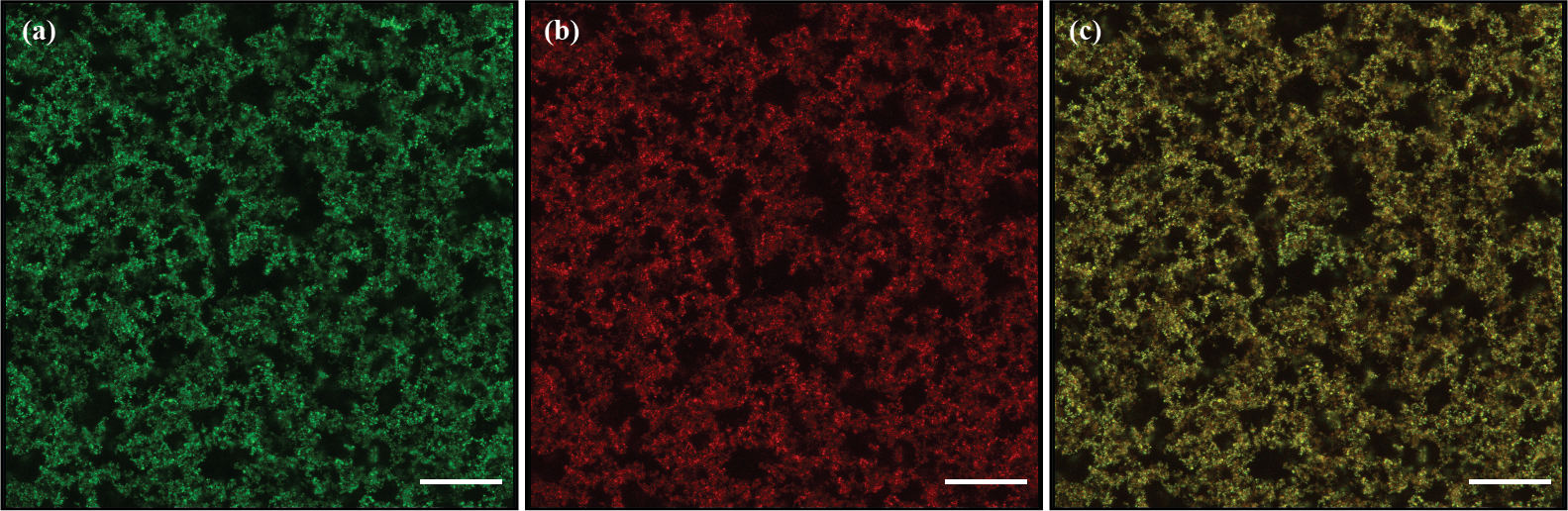}
\caption{\label{supp_confocal} Confocal fluorescent micrographs of the strong particles (a), and the weak particles (b) in a binary gel network for a mixing proportion $\chi=0.5$ and a total particle volume fraction $\phi=1.0\%$ at $T=30\si{\celsius}$. (c) Merged image of (a) and (b). Both colors are found uniformly throughout the image, indicating a homogeneous composition of the mixture in the gelled state. The three-dimensional projection is performed over $31.5\;\si{\micro\meter}$ in the direction of the optical axis for all the images. Scale bars correspond to $30\;\si{\micro\meter}$.}
\end{figure*}

\begin{figure*}[t]
\setlength{\abovecaptionskip}{-30pt}
\hspace*{-0.06cm}\includegraphics[scale=0.115]{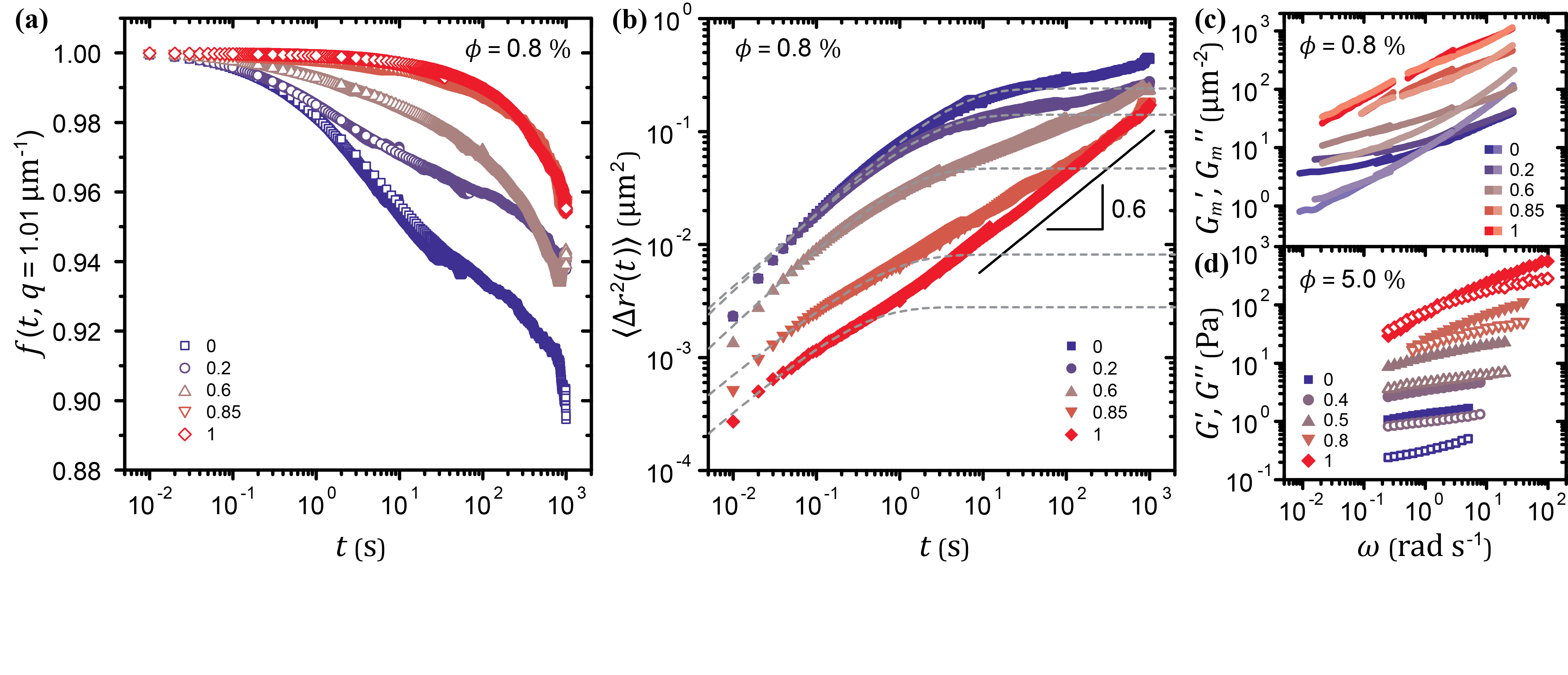}
\caption{\label{fig1}(a) Normalized intermediate scattering function $f$ for different mixing proportions $\chi$ at wave vector $q=1.01\;\si{\per\micro\meter}$ and total particle volume fraction $\phi=0.8\%$. (b) Mean squared displacement (MSD) $\left<{\Delta}r^2(t)\right>$ of clusters and corresponding stretched exponential fits at early times (dashed lines) for $\phi=0.8\%$. (c) Modified storage modulus ${G_m}'$ (darker lines) and modified loss modulus ${G_m}''$ (lighter lines) as a function of frequency $\omega$ for $\phi=0.8\%$, obtained from the generalized Stokes-Einstein relation. (d) Storage modulus $G'$ (filled) and loss modulus $G''$ (open) measured in rheometry for $\phi=5.0\%$. Legends indicate $\chi$.}
\end{figure*}

The gels for any mixing proportion $\chi$ exhibit the characteristics of kinetically arrested networks of fractal clusters. The temporal change in the relaxation time of the aggregating weak particles alone ($\chi=0$) shows the kinetics of diffusion-limited cluster aggregation (DLCA) \cite{Meakin1983,Weitz1984a,Weitz1984,vanDongen1985,Cho2020}. Additionally, we find a $\chi$-independent fractal dimension $d_{f}=1.8\pm0.1$ of the clusters, a structural characteristic of DLCA gels \cite{Weitz1984a}, based on the static structure factor $S(q)$ obtained from the micrographs \cite{Carpineti1992,Lu2012,Giavazzi2014,Cho2020}. Given that the two types of particles of comparable radii are homogeneously mixed before gelation, diffusion-limited aggregation upon the rapid temperature increase leads to a percolating network of uniformly sized clusters, whose radius $R_{c}\;{\approx}\;a\phi^{-1/(3-d_f)}$ depends primarily on $\phi$ \cite{DelGado2016}, while $\chi$ determines the effective attraction strength. The uniformity in composition of the gel networks is confirmed with confocal fluorescent micrographs (Leica, TCL SP8) of the two types of particles labeled with different colors (Pyrromethene 546 and 650, Exciton) as displayed in Fig.~\ref{supp_confocal} \cite{Immink2019}. We thus extract the dynamics of the clusters in the form of the ensemble-averaged MSD $\left<{\Delta}r^2(t)\right>$ from $f(t,q)=\exp\left[-q^2\left<{\Delta}r^2(t)\right>/4\right]$ \cite{Pusey2002,Bayles2017,Edera2017}, as shown in Fig.~\ref{fig1}(a,b) for $\phi=0.8\%$. In our experiments, DDM captures the motion of interacting clusters under partially coherent illumination, which strictly renders this relation inapplicable \cite{Pusey2002,Giavazzi2009,Giavazzi2014}. Nonetheless, the approximation has been shown to effectively capture the $q$-independent dynamics dominated by the clusters \cite{Krall1997,Krall1998,Cho2020}, and we observe that the resultant $\left<{\Delta}r^2(t)\right>$ is indeed independent of $q$ over multiple length scales for our samples (see Appendix~\ref{ddm_detail}). All $\left<{\Delta}r^2(t)\right>$ reported here are averaged over 19 consecutive $q$ in the range of $0.70\;{\leq}\;q\;{\leq}\;1.51\;\si{\per\micro\meter}$ that lies in the cluster-dominated regime for all $\phi$ and $\chi$ explored. \par

The fluctuations of the clusters are known to constitute the floppiest mode of the gel dynamics, which governs the macroscopic elasticity of the gels \cite{Kantor1984,Shih1990,deRooij1994,Krall1998}. We here show that the cluster motion also manifests the macroscopic \textit{visco}elasticity. Treating the clusters as tracer particles, we convert $\left<{\Delta}r^2(t)\right>$ to the viscoelastic moduli through the creep compliance $J(t)$, using the generalized Stokes-Einstein relation in the time domain \cite{Mason1995,Xu1998,Squires2010}:
\begin{equation}
J(t)=\frac{3{\pi}r_{t}}{2k_{B}T}\left<{\Delta}r^2(t)\right>, \quad  G^*(\omega)=\frac{1}{i\omega\hat{J}(\omega)}, \label{GSE}
\end{equation}
where $r_t$ denotes the tracer radius, $k_B$ the Boltzmann constant, $G^*(\omega)=G'(\omega)+iG''(\omega)$ the complex modulus as a function of the frequency $\omega$, and $\hat{J}(\omega)$ the Fourier-transformed $J(t)$. Since Eq.~\ref{GSE} assumes spherical probes, $r_t$ of the fractal clusters cannot be unambiguously determined. Yet, the cluster radius $R_c$ and the fractal dimension $d_f$, both independent of $\chi$, are expected to determine the effective $r_t$. Hence, for the binary gels of different $\chi$ at $\phi=0.8\%$, we report the modified modulus ${G_{m}}^*(\omega)=3{\pi}r_tG^{*}(\omega)/(2k_{B}T)=1/\left[i\omega \left<{\Delta}\hat{r}^2(\omega)\right>\right]$, where $\left<{\Delta}\hat{r}^2(\omega)\right>$ denotes the Fourier-transformed MSD, via the direct conversion \cite{Evans2009} after smoothing, shown in Fig.~\ref{fig1}(c). The smoothing filters the noise significantly, without distorting the general trends in the data, as illustrated in Appendix~\ref{smoothing}. The lower values and the stronger time dependence at late $t$ of $\left<{\Delta}r^2(t)\right>$ with increasing $\chi$ in Fig.~\ref{fig1}(b) translate into the higher moduli and the stronger frequency dependence at low $\omega$, respectively, in Fig.~\ref{fig1}(c). The identical trends are observed in the moduli measured with rheometry at $\phi=5.0\%$, as displayed in Fig.~\ref{fig1}(d). Moreover, the crossover between the storage and the loss moduli at $\omega\approx1.2\;\si{\radian\per\s}$ for $\chi=1$  is captured at both $\phi$, further indicating that the cluster dynamics represents the macroscopic viscoelasticity. \par

The dominance of ${G_m}''$ over ${G_m}'$ at high frequencies for $\phi=0.8\%$, or equivalently the steep increase in $\left<{\Delta}r^2(t)\right>$ at early times, arises from the competition between the localized elastic response within each cluster and the solvent viscosity. The overdamped dynamics can be modeled as a stretched exponential function $\left<{\Delta}{r_l}^2(t)\right>=\delta^2\left\{1-\exp\left[-\left(t/\tau\right)^p\right]\right\}$,  where $\delta^2$ denotes the maximum localized MSD, $\tau$ the relaxation time, and $p=0.66\pm0.03$ the stretching exponent, as shown by the fits in Fig.~\ref{fig1}(b) \cite{Krall1998,Cho2020}. The details of the fitting procedures are described in Appendix~\ref{fit_method}. The relaxation time $\tau$ is set by the spring constant of the stress-bearing backbones of the clusters \cite{Kantor1984,Krall1998}, and thus $\tau$ decreases with the proportion of the strong particles $\chi$. Likewise, as $\phi$ increases, the clusters become smaller and stiffer \cite{Kantor1984,Shih1990}, which in turn reduces $\tau$, causing the dominance of $G''$ at high frequencies to be experimentally inaccessible at the higher $\phi=5.0\%$, as displayed in Fig.~\ref{fig1}(d). \par

\begin{figure}[b]
\setlength{\abovecaptionskip}{-15pt}
\hspace*{-0.06cm}\includegraphics[scale=0.125]{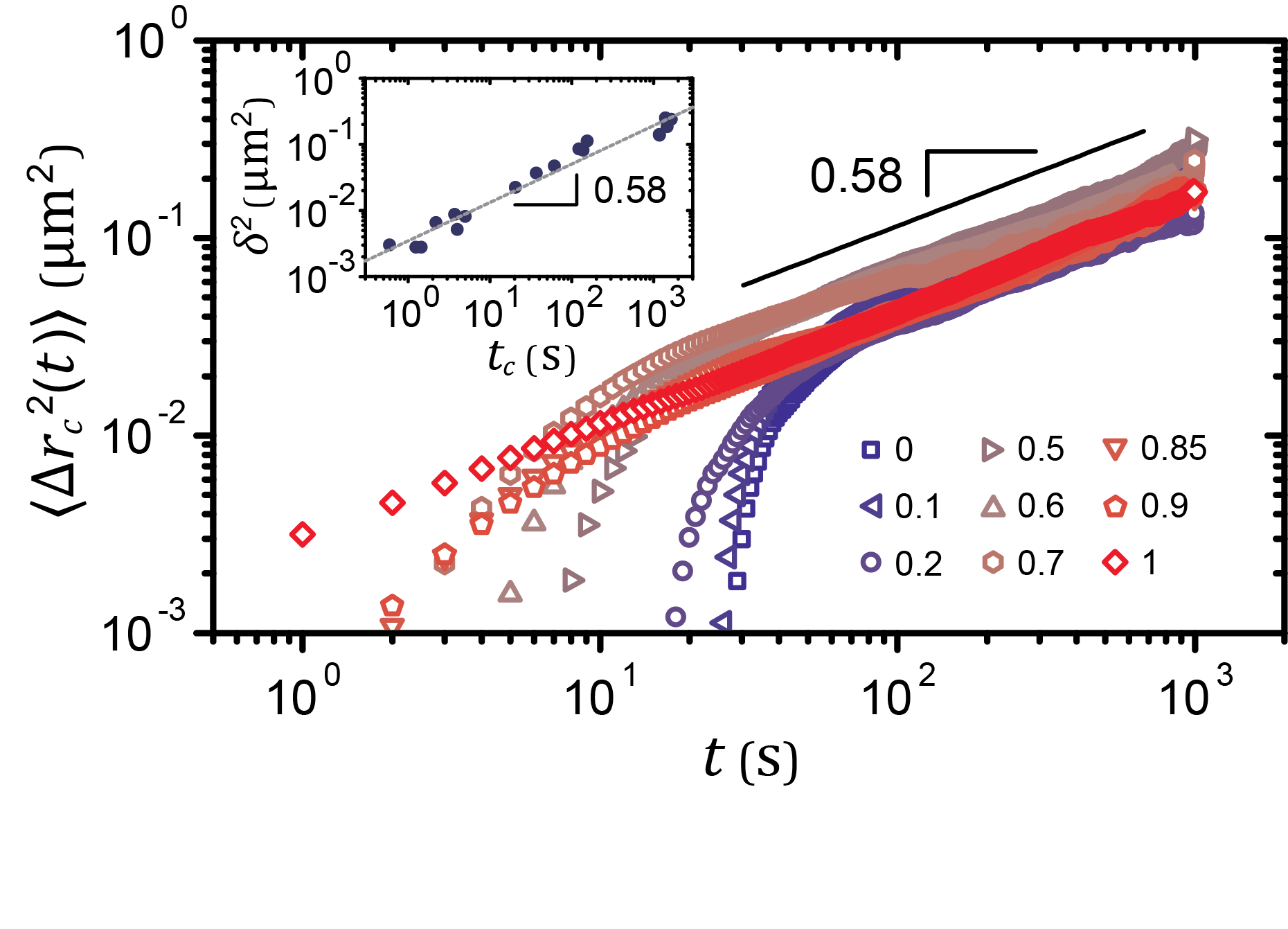}
\caption{\label{fig2} Collective part $\left<{\Delta}{r_c}^2(t)\right>$ of the MSD for different mixing proportions $\chi$ at $\phi=0.8\%$. $\left<{\Delta}{r_c}^2(t)\right>$ cannot be accurately determined at early times when the localized part is dominant, which causes the apparent, unphysical rapid increase at early $t$. Inset: Maximum MSD ${\delta}^2$ of the localized part as a function of the transition time $t_c$ into the collective-mode-dominant regime. The dotted line denotes the power-law fit.}
\end{figure}

The deviation of $\left<{\Delta}r^2(t)\right>$ at late $t$ from the plateaus~$\delta^2$ shown in Fig.~\ref{fig1}(b), however, corroborates that the clusters do not remain localized, as they gradually fluctuate more cooperatively. To decouple the collective portion $\left<{\Delta}{r_c}^2(t)\right>$ of the MSD from the localized one $\left<{\Delta}{r_l}^2(t)\right>$, we assume that the two modes are additive and independent. This assumption allows us to quantify the collective fluctuations as $\left<{\Delta}{r_c}^2(t)\right>=\left<{\Delta}r^2(t)\right>-\left<{\Delta}{r_l}^2(t)\right>$, as shown in Fig.~\ref{fig2} for $\phi=0.8\%$, which unveils a $\chi$-independent power law with an exponent $0.58\pm0.03$ at late times. This $\chi$ independence is further demonstrated by the transition time $t_c$ from the early localized response to the late subdiffusion. We estimate $t_c$ by solving $\delta^2=K{t_c}^{\nu}$, where $K$ and $\nu$ are obtained from power law fits to the late-time $\left<{\Delta}{r_c}^2(t)\right>$. The estimates for all $\chi$ exhibit the scaling $\delta^2\;{\sim}\;{t_c}^{0.58\pm0.03}$, as shown in the inset of Fig.~\ref{fig2}, consistent with the exponent of $\left<{\Delta}{r_c}^2(t)\right>$. \par

The $\chi$ independence of the late-time subdiffusion indicates that the hydrodynamic coupling or the steric hindrance, rather than the attractive interactions, underlies the cooperative fluctuations of the network. In much weaker and denser gels, hydrodynamic interactions among individual particles can significantly affect the viscoelasticity \cite{Varga2018}. In our gels, however, the clusters are fractal and multiply-bonded, which renders the dominance of the hydrodynamic coupling unlikely. The connectivity of the gel network, in fact, requires that each cluster, surrounded by its neighbors, is sterically constrained in a cage of approximately its own size, which is largely $\chi$-independent. Within the duration of image acquisition, we find that $\left<{\Delta}r^2(t)\right>\ll{R_{c}}^2$, as shown in Fig.~\ref{fig1}(b), where the DLCA cluster radius $R_{c}\;{\approx}\;5.0 - 6.5\;\si{\micro\meter}$ at $\phi=0.8\%$. The small MSD suggests that the subdiffusion arises within the cages, as confirmed by the preservation of the network configuration in the real-space image sequences \cite{Note1}. Thus, this collective dynamics contrasts with the aging dynamics accompanying large-scale structural rearrangements \cite{Cipelletti2000,Bouzid2017,Chaudhuri2017}, which we partially observe in our systems at even later times as displayed in Appendix~\ref{aging}. The cooperative fluctuations mediated by the steric hindrance bear resemblance to the motion of stable colloidal particles confined in a one-dimensional channel that prevents them from passing one another. In such single-file systems, the excluded passage maintains the configuration of the particles, giving rise to their subdiffusion with an exponent slightly greater than or equal to $0.5$ \cite{Wei2000,Kollmann2003,Lutz2004,Taloni2008,Lizana2010,Taloni2017,Euan-Diaz2012}. Moreover, many-body interactions of particles under mutual confinement have been modeled as viscoelastic forces applied to the individual constituents through the generalized Langevin equation \cite{Taloni2008,Lizana2010,Plyukhin2019} from which Eq.~(\ref{GSE}) can be derived \cite{Mason1995,Squires2010}. This formulation establishes an explicit link between the steric hindrance among the clusters embedded in a network and their viscoelastic subdiffusion \cite{Goychuk2009}. \par

\begin{figure}[t]
\setlength{\abovecaptionskip}{-15pt}
\hspace*{-0.06cm}\includegraphics[scale=0.125]{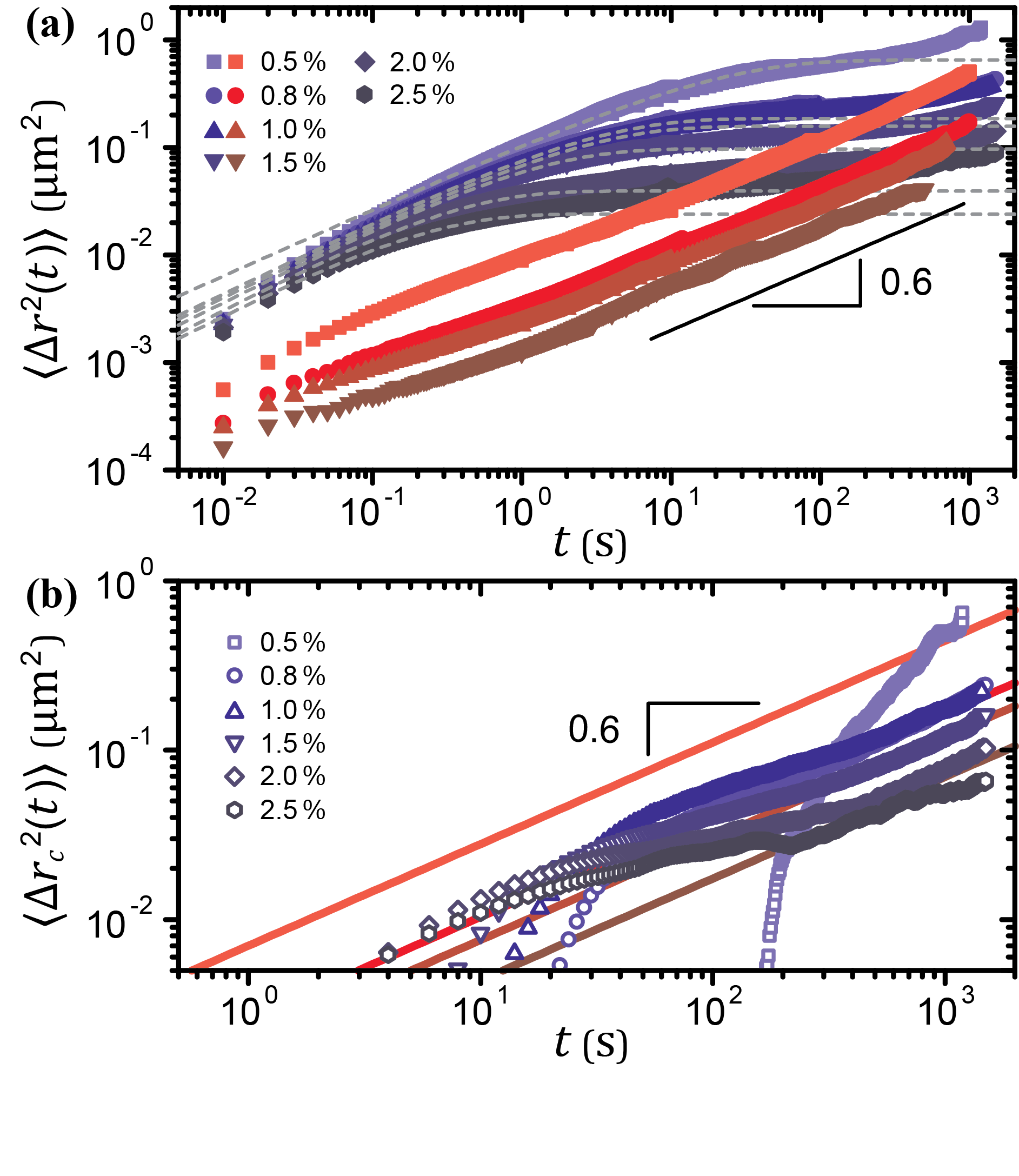}
\caption{\label{fig3} (a) Mean squared displacement $\left<{\Delta}{r}^2(t)\right>$ for the mixing proportions $\chi=0$ (blue) and $\chi=1$ (red) at different $\phi$, and the stretched exponential fits for $\chi=0$ (dashed lines) at early times. The MSD for $\chi=1$ and $\phi>1.5\%$ cannot be reliably obtained because of vanishingly small fluctuations. (b) Collective part $\left<{\Delta}{r_c}^2(t,\chi=0)\right>$ of the MSD at different $\phi$. Solid lines with slope 0.6 correspond to the locations of the late-time $\left<{\Delta}{r}^2(t,\chi=1)\right>$ for the four $\phi$ shown in (a).}
\end{figure}

The collective subdiffusion emerges also at other particle volume fractions, as the cooperativity of the confined clusters due to the steric hindrance remains in effect. For $\chi=1$, where the late-time dynamics is dominated by the collective mode, we indeed find $\left<{\Delta}{r}^2(t)\right>\ \sim\ t^{0.6}$ for $0.5\%\leq\phi\leq1.5\%$, as shown in Fig.~\ref{fig3}(a). For $\chi=0$, although the initial elastic response eclipses the collective mode over most of the accessible times at $\phi=0.5\%$, $\left<{\Delta}{r_c}^2(t)\right>$ at higher $\phi$ unveils the subdiffusion at late times, as shown in Fig.~\ref{fig3}(b). We note that as the onset of the subdiffusion in denser systems is delayed, a transitional dynamics appears at intermediate times, which renders the power-law exponent of 0.6 inaccessible at $\phi=2.5\%$. This slower occurrence of the long-time behavior $\left<{\Delta}{r_c}^2(t)\right>=Kt^{0.6}$ at higher $\phi$ could be understood as a result of the decreased prefactor $K$ due to the greater confinement among more closely packed neighboring clusters, whose size decreases with $\phi$. \par

\begin{figure}[t]
\setlength{\abovecaptionskip}{-0pt}
\hspace*{-0.06cm}\includegraphics[scale=0.125]{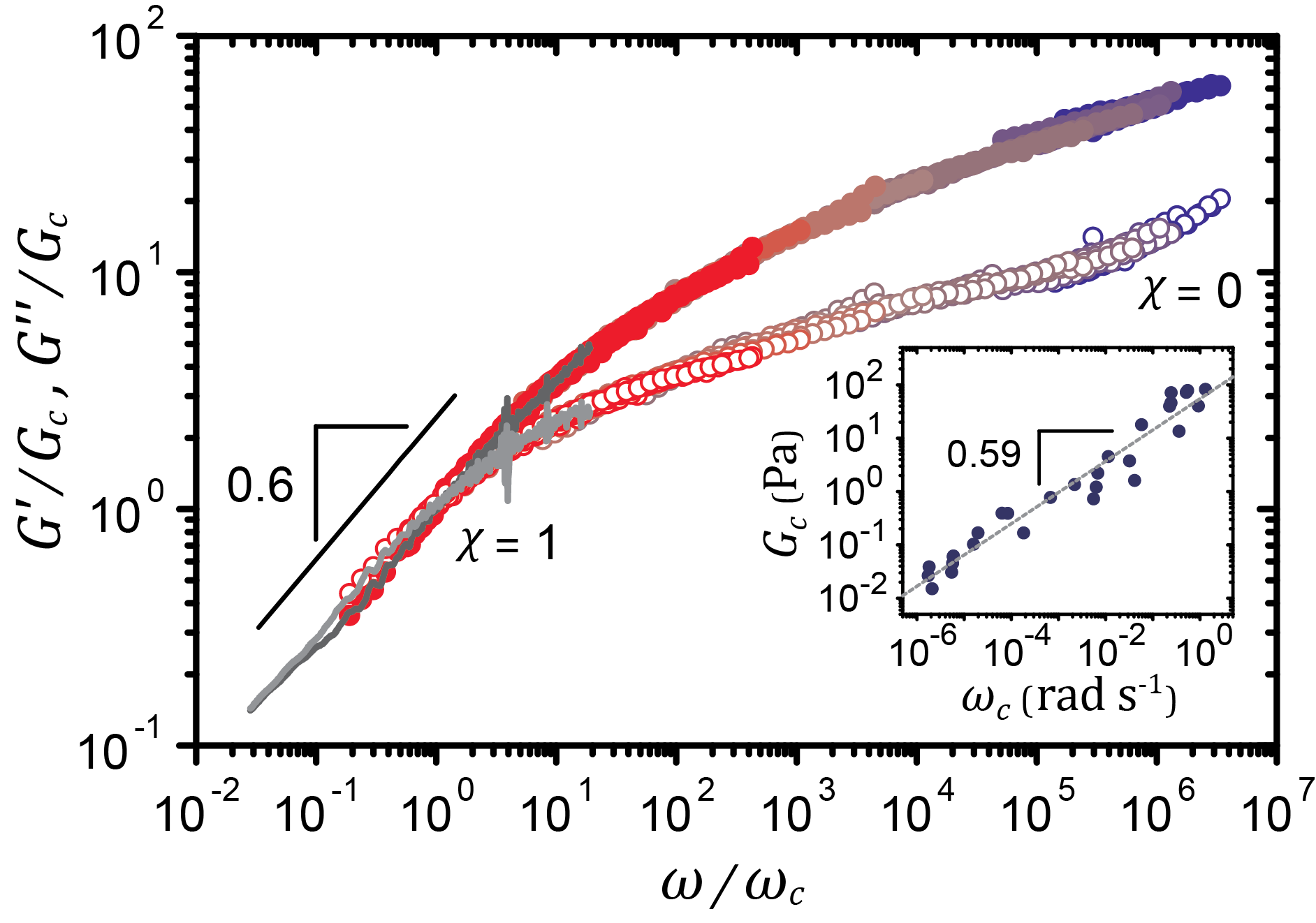}
\caption{\label{fig4} Storage modulus $G'$ (filled) and loss modulus $G''$ (open) scaled with the crossover frequency $\omega_{c}$ in the abscissa and the crossover modulus $G_c$ in the ordinate for different mixing proportions $\chi$ at $\phi=5.0\%$. The dark ($G'$) and light ($G''$) gray lines, which display a power law with an exponent 0.6 at low $\omega/{\omega_c}$, represent the converted moduli from the creep compliance for $\chi=1$ \cite{Evans2009}. Inset: $G_c$ as a function of $\omega_{c}$. The dotted line denotes the power-law fit.}
\end{figure}

We show that the delayed onset of the collective subdiffusion at higher $\phi$, nevertheless, controls the systematic change in the dynamics with $\chi$, and thus enables us to tune the viscoelastic moduli of the binary gels over an even broader range of frequencies. The full spectrum of the variation in $G'$ and $G''$ with $\chi$ at $\phi=5.0\%$, displayed in Fig.~\ref{fig1}(d), can be obtained by constructing a master curve, as shown in Fig.~\ref{fig4}. In order of decreasing $\chi$, we scale each set of the moduli with respect to the crossover frequency $\omega_c$ in the abscissa and the crossover modulus $G_c$ in the ordinate. After finding $\omega_c$ first by scaling the loss tangent $G''/G'$, we identify the corresponding $G_c$ that leads to the master curve, which we verify with the Kramers-Kronig relations \cite{Parot2007,Rouleau2013}, as illustrated in Appendix~\ref{KKrelation}. The resulting scaling parameters exhibit a power law $G_c\: \sim \: {\omega_c}^{0.59\pm0.03}$, as displayed in the inset of Fig.~\ref{fig4}. Under the assumption that the gels of different $\chi$ at $\phi=5.0\%$ exhibit identical intermediate dynamics, a characteristic transition time $t_c$ can be consistently defined by $\delta^2=K{t_c}^{\nu}$. The scaling relation between $G_c$ and $\omega_c$ can then be translated into $\delta^2\: \sim \: {t_c}^{0.59\pm0.03}$ by Eq.~(\ref{GSE}), which indicates that the location of the elastic plateau with respect to the $\chi$-independent late-time subdiffusion determines the linear viscoelasticity over six decades of timescales. We highlight that, corresponding to $\left<{\Delta}{r}^2(t,\chi=1)\right>$ at late times, $G'/G_c$ and $G''/G_c$ at low frequencies obey a power law with an exponent 0.6, further justifying our use of Eq.~(\ref{GSE}). \par

\section{Conclusions}
Using two complementary techniques, DDM and rheometry, we demonstrate that the microscopic fluctuations of the clusters in dilute colloidal gels can be decomposed into a localized elastic mode and a collective viscoelastic mode, whose interplay determines the macroscopic viscoelasticity. We suggest that the steric hindrance among the clusters of the percolating network governs the long-time collective dynamics, unlike the early-time localized dynamics determined by the mean attraction strength between the particles. This difference in the underlying mechanisms of the two modes allows us to tune the viscoelastic moduli of the binary gels by changing the mixing proportion, while keeping $\phi$ constant. When combined with a purely elastic response, a mode of fluctuation with a fixed power-law exponent of 0.6 can yield diverse viscoelastic spectra. Thus, the characterization of the dynamics as the superposition of the two modes can enhance the microscopic interpretations of different exponents in power-law rheology models, often used to describe the linear viscoelasticity of a broad range of soft materials \cite{Jaishankar2013,Aime2018b,Bonfanti2020}. Furthermore, we underscore that simply increasing the stiffness of a gel can transform its viscoelastic spectrum, and even generate, rather counterintuitively, the apparent fluidlike behavior $G''>G'$, as shown in Fig.~\ref{fig4}. This versatility of the binary gels elucidates that the many-body dynamics of the constituents in complex fluids can be harnessed to methodically tune their mechanical properties. Specifically, the concurrence of the two modes in our gels originates from the structural hierarchy in which the network is composed of multiply connected building blocks, \textit{i.e.}, clusters, each of which consists of aggregated particles. Such hierarchy, in fact, pervades a number of natural systems, such as protein gels \cite{Gibaud2013} and multicellular networks \cite{Leggett2019}, in addition to various man-made materials, from traditional cement \cite{Jennings2000}, to novel functionalized gels \cite{Fan2014,Diba2017,Nair2019,Xiong2019}. Our results, therefore, may provide general design guidelines for complex soft materials composed of aggregated nanoparticles. \par

\begin{acknowledgments}
We thank Roberto Cerbino, Emanuela Del Gado, Gareth H. McKinley, James W. Swan, and Veronique Trappe for helpful discussions. We also thank Ming Guo for providing us with access to the confocal microscope. We acknowledge support from the MIT Research Support Committee and Kwanjeong Educational Foundation, Awards No. 16AmB02M and No. 18AmB59D. 
\end{acknowledgments}

\appendix
\section{Calculation of the intermediate scattering function $f\left(q,t\right)$ and the mean squared displacement $\left<{\Delta}{r}^2(t)\right>$}
\label{ddm_detail}
The image structure function $D(q,t)$, where $q$ denotes the wave vector and $t$ the delay time, computed with the DDM algorithm can be expressed as
\begin{equation}
D(q,t)=A(q)\left[1-f(q,t)\right]+B(q), \label{eq1}
\end{equation}
where $A(q)$ is determined by optical properties of the microscope and static information about the sample, and $B(q)$ represents the level of the camera noise \cite{Giavazzi2014}. We find the normalized intermediate scattering function $f(q,t)$ by first determining $A(q)$ and $B(q)$. The value of $B(q)$ is independent of $q$ if the detection noise of the camera is uncorrelated in space and time \cite{Giavazzi2017}. Indeed we observe that $D(q,t)$  becomes independent of both $q$ and $t$ in the highest-$q$ domain accessible ($q>28\;\si{\per\micro\meter}$ for the $60\times$ objective), where $A(q)$ approaches zero. We calculate the mean of $D(q,t)$ in this domain and equate the resulting value to $B(q)$. \par

In linear space invariant imaging, where the sample density field is linearly mapped onto the image intensity field, the ensemble-averaged squared modulus of Fourier-transformed images can be expressed as
\begin{equation}
\left<{\abs{\hat{i}(q)}}^{2}\right>_{E} \simeq \frac{A(q)}{2}+\frac{B(q)}{2}, \label{eq2}
\end{equation}
provided that the non-ideal contributions arising from imperfections, such as scratches, stains, or dust particles, along the optical path are negligible compared to the signals of the scattered light from the sample \cite{Giavazzi2014}. Hence we obtain $A(q)$ from $\left<{\abs{\hat{i}(q)}}^{2}\right>_{E}$ and $B(q)$. \par

\begin{figure}[b]
\setlength{\abovecaptionskip}{-15pt}
\hspace*{-0.03cm}\includegraphics[scale=0.11]{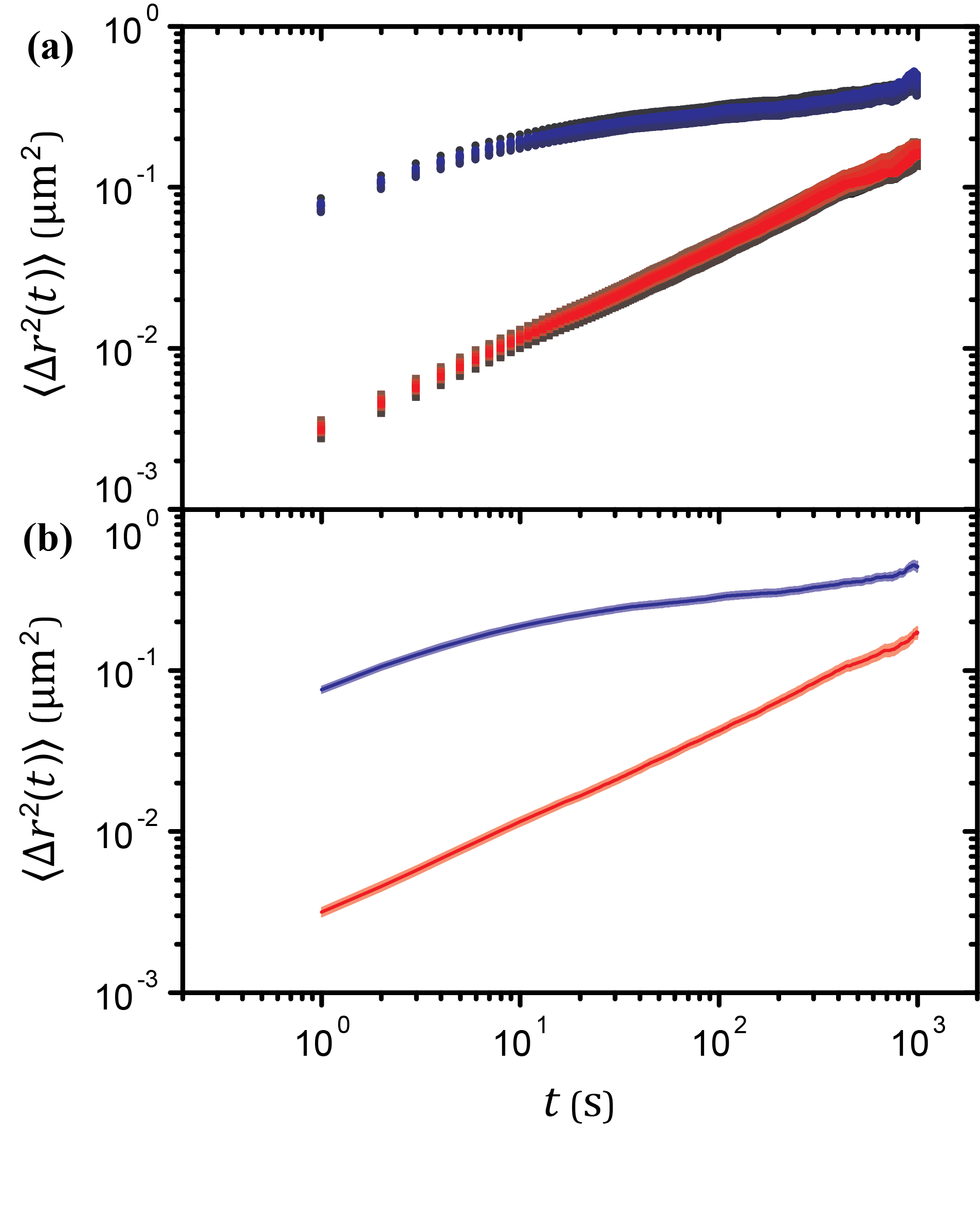}
\caption{\label{supp_MSD} (a) Mean squared displacement $\left<{\Delta}{r}^2(t)\right>$ of the strong ($\chi=1$, red squares) and the weak ($\chi=0$, blue circles) gels at $\phi=0.8\%$ calculated with $f(q,t)=\exp\left[-q^2\left<{\Delta}r^2(t)\right>/4\right]$ at 19 consecutive wave vectors $q$. Brighter data sets represent higher $q$. (b) Corresponding $\left<{\Delta}{r}^2(t)\right>$ averaged over the 19 wave vectors (darker lines) and the standard deviations (lighter areas).}
\end{figure}

We calculate the mean squared displacement (MSD) of the clusters $\left<{\Delta}{r}^2(t)\right>$ by using the relation $f(q,t)=\exp\left[-q^2\left<{\Delta}r^2(t)\right>/4\right]$ \cite{Pusey2002,Bayles2017,Edera2017}, which assumes scattering from identical, noninteracting particles \cite{Pusey2002}. In applying this relation to extract the MSD of the interacting clusters under partially coherent illumination, we assume the cluster-dominated fluctuations, where the cluster motion dominates $f(q,t)$ over multiple length scales \cite{Krall1998,Cho2020}. We indeed find $q$ independence of the resulting $\left<{\Delta}{r}^2(t)\right>$, and report the averaged MSD over a range of the wave vectors $0.70\;{\leq}\;q\;{\leq}\;1.51\;\si{\per\micro\meter}$, as shown in Fig.~\ref{supp_MSD}(a,b). \par

\section{Conversion of the mean squared displacement $\left<{\Delta}{r}^2(t)\right>$ into the modified viscoelastic moduli ${G_m}'(\omega)$, ${G_m}''(\omega)$}
\label{smoothing}
\begin{figure}[t]
\setlength{\abovecaptionskip}{-25pt}
\hspace*{-0.03cm}\includegraphics[scale=0.11]{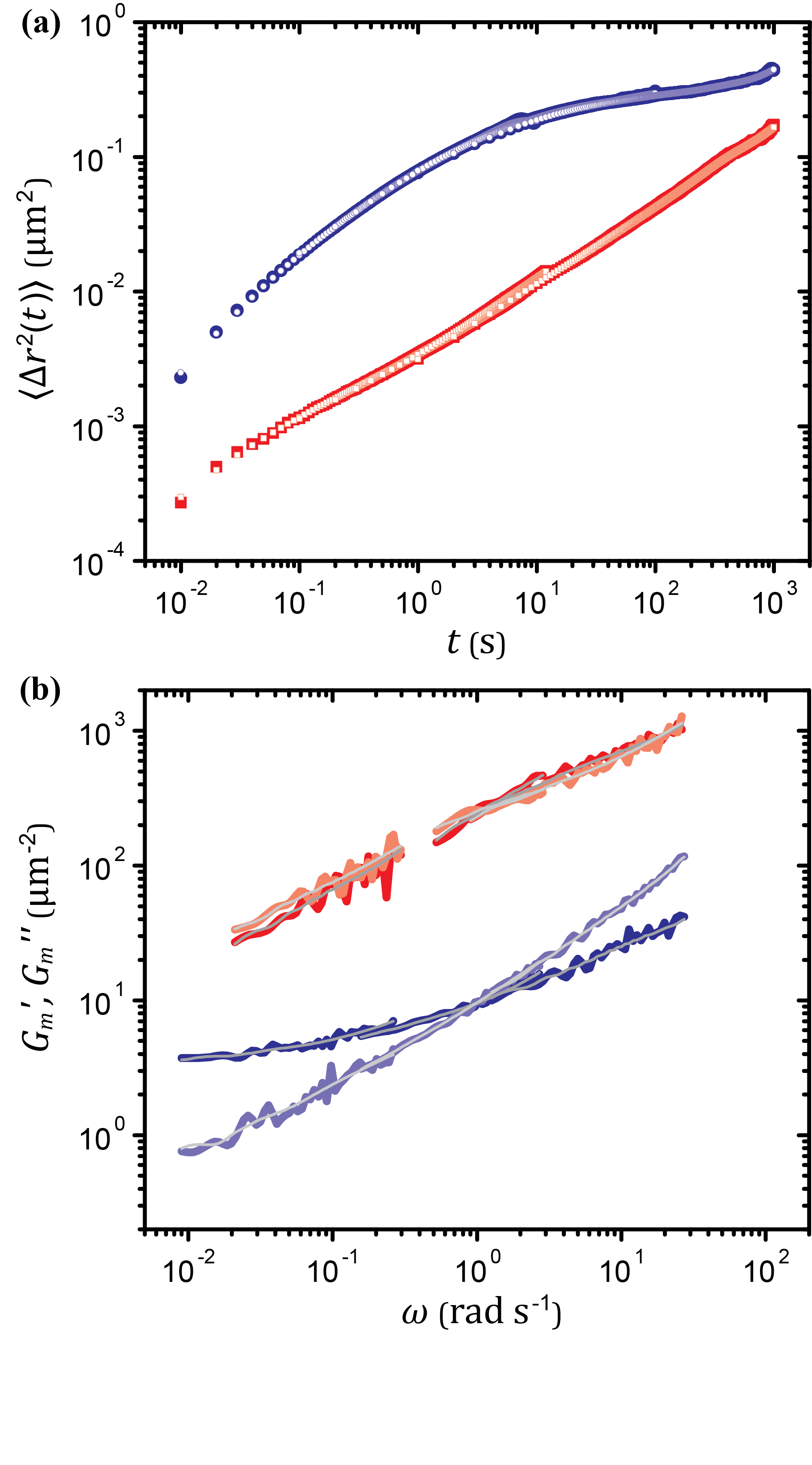}
\caption{\label{supp_smoothing} (a) Mean squared displacement $\left<{\Delta}{r}^2(t)\right>$ of the strong ($\chi=1$, red filled squares) and the weak ($\chi=0$, blue filled circles) gels at $\phi=0.8\%$ and the corresponding smoothed $\left<{\Delta}{r}^2(t)\right>$ (open, smaller symbols). (b) Modified viscoelastic moduli ${G_m}'$ (darker lines) and ${G_m}''$ (lighter lines) converted from $\left<{\Delta}{r}^2(t)\right>$ in (a). The thicker colored lines result from the raw $\left<{\Delta}{r}^2(t)\right>$ and the thinner gray lines result from the smoothed $\left<{\Delta}{r}^2(t)\right>$.}
\end{figure}

We convert the mean squared displacement $\left<{\Delta}{r}^2(t)\right>$ into the modified viscoelastic moduli 
\begin{equation}
{G_m}'(\omega)=\frac{3{\pi}r_t}{2{k_B}T}G'(\omega),\quad {G_m}''(\omega)=\frac{3{\pi}r_t}{2{k_B}T}G''(\omega), \label{eq3}
\end{equation}
where $r_t$ denotes the tracer radius and $k_B$ the Boltzmann constant, via the generalized Stokes-Einstein relation $J(t)=3{\pi}r_{t}\left<{\Delta}r^2(t)\right>/\left(2k_{B}T\right)$ and $G^*(\omega)=G'(\omega)+iG''(\omega)=1/\left[i\omega\hat{J}(\omega)\right]$, where $J(t)$ denotes the creep compliance and $\hat{J}(\omega)$ its Fourier transform, after smoothing $\left<{\Delta}{r}^2(t)\right>$ with the function \textit{csaps} (smoothing parameter $= 0.95$) in MATLAB. The cubic smoothing spline interpolation of the data sets minimally alters the general behavior of $\left<{\Delta}{r}^2(t)\right>$ as shown in Fig.~\ref{supp_smoothing}(a), while effectively reducing the level of noise in the converted moduli as displayed in Fig.~\ref{supp_smoothing}(b). \par

\section{Fitting method for stretched exponential functions}
\label{fit_method}
We use \textit{lsqcurvefit} in MATLAB to fit stretched exponential functions to the mean squared displacement (MSD) $\left<{\Delta}{r}^2(t)\right>$ at early times, to obtain the localized part of the MSD $\left<{\Delta}{r_l}^2(t)\right>=\delta^2\left\{1-\exp\left[-\left(t/\tau\right)^p\right]\right\}$, where the maximum localized MSD $\delta^2$, the relaxation time $\tau$, and the stretching exponent $p$ are fitting parameters. The absence of well-defined plateaus in $\left<{\Delta}{r}^2(t)\right>$ calculated from the normalized intermediate scattering function requires a systematic approach to obtaining the fitting parameters \cite{Cho2020}. We plot $\left<{\Delta}{r}^2(t)\right>$ in linear-log scales to estimate the time window of fitting $[0, t^*]$ by ensuring that a point of inflection is captured in the domain. Such a point of inflection indicates that $\tau\;{\lesssim}\;t^*$, which enables us to properly identify $\delta^2$. To improve the reliability of the fitting values of $\tau$ and $p$, we then linearize the stretched exponential function by plotting $\log \left\{ -\log \left[ 1-\frac{\left<r^2(t)\right>}{\delta^2} \right] \right\}$ as a function of $\log(t)$, such that the slope is equal to $p$ and the $y$-intercept is equal to $-p \log(\tau)$. The resulting plot typically deviates from the linear behavior at later $\log(t)$ by showing a gradually decreasing slope. By performing linear regression to the data in the linear regime, we re-evaluate $\tau$ and $p$. We finally use these values to plot the stretched exponential function and check the quality of the fit. \par

\clearpage
\section{Nonequilibrium aging dynamics at later times}
\label{aging}
\begin{figure}[b]
\setlength{\abovecaptionskip}{-15pt}
\hspace*{-0.03cm}\includegraphics[scale=0.11]{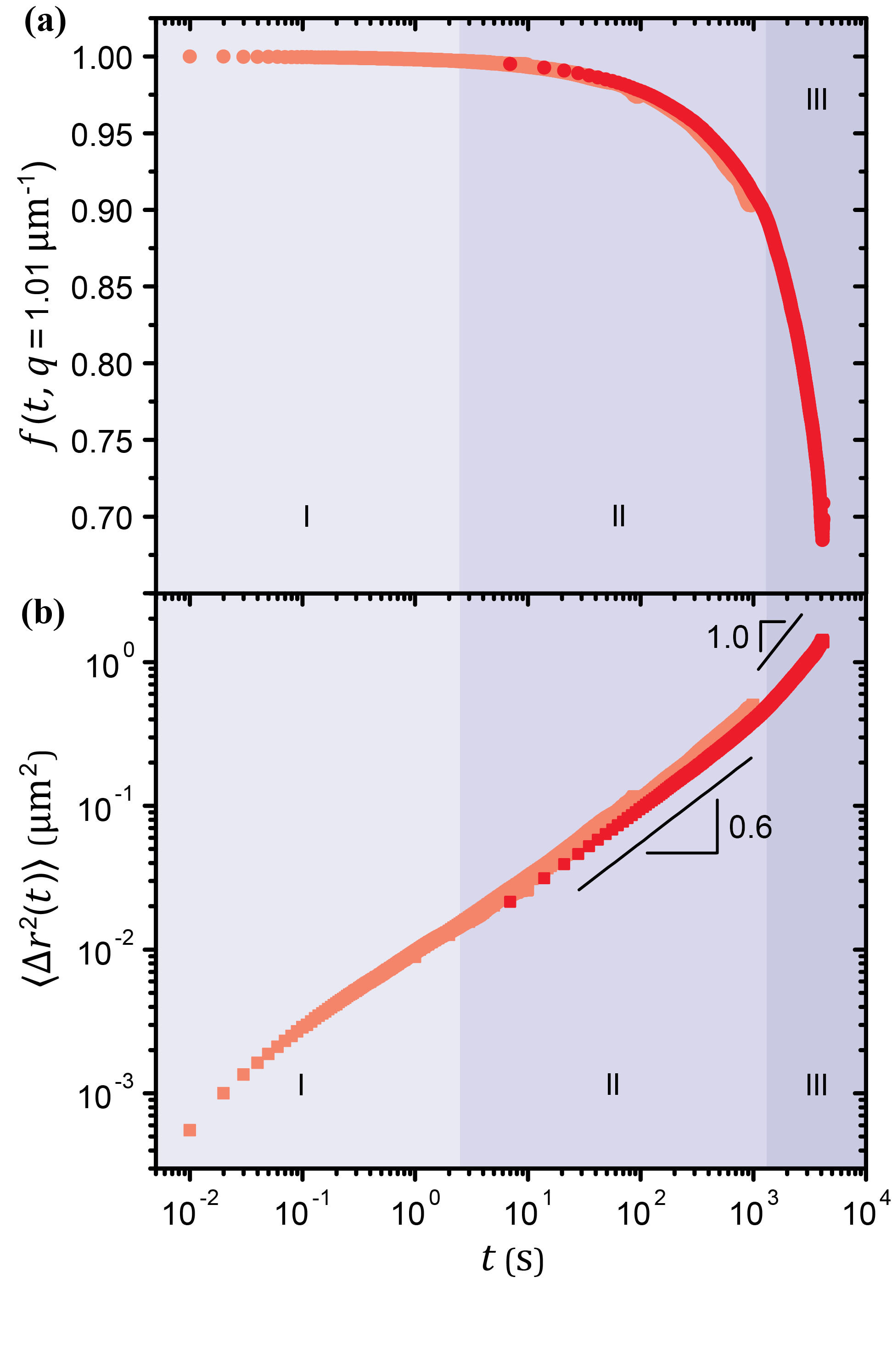}
\caption{\label{supp_aging} (a) Normalized intermediate scattering function $f(t)$ at wave vector $q=1.01\;\si{\per\micro\meter}$ and (b) mean squared displacement $\left<{\Delta}{r}^2(t)\right>$ of the gel for $\phi=0.5\%$ and $\chi=1$, calculated from image sequences with four different interval times. The data points in the darker red represent the results obtained from the image sequence with the largest interval time ($7\;\si{\s}$). Three temporal regimes of different dominant modes are found: the localized elastic mode (I), the collective viscoelastic mode (II), and the aging dynamics (III).}
\end{figure}

Using image sequences acquired over longer periods than those reported in the main text, we can have access to the onset of the aging dynamics, as shown in Fig.~\ref{supp_aging} for $\phi=0.5\%$ and $\chi=1$. Although the aging dynamics cannot be readily differentiated from the two quasiequilibrium modes in the normalized intermediate scattering function $f(t,q=1.01\;\si{\per\micro\meter})$, the steeper increase in the mean squared displacement $\left<{\Delta}{r}^2(t)\right>$ displayed at latest times marks the aging dynamics, which may be diffusive or superdiffusive \cite{Cipelletti2000,Bouzid2017,Chaudhuri2017}. \par

\newpage
\section{Application of the Kramers-Kronig relations to the master curve of the viscoelastic moduli}
\label{KKrelation}
\begin{figure}[b]
\setlength{\abovecaptionskip}{10pt}
\hspace*{-0.03cm}\includegraphics[scale=0.11]{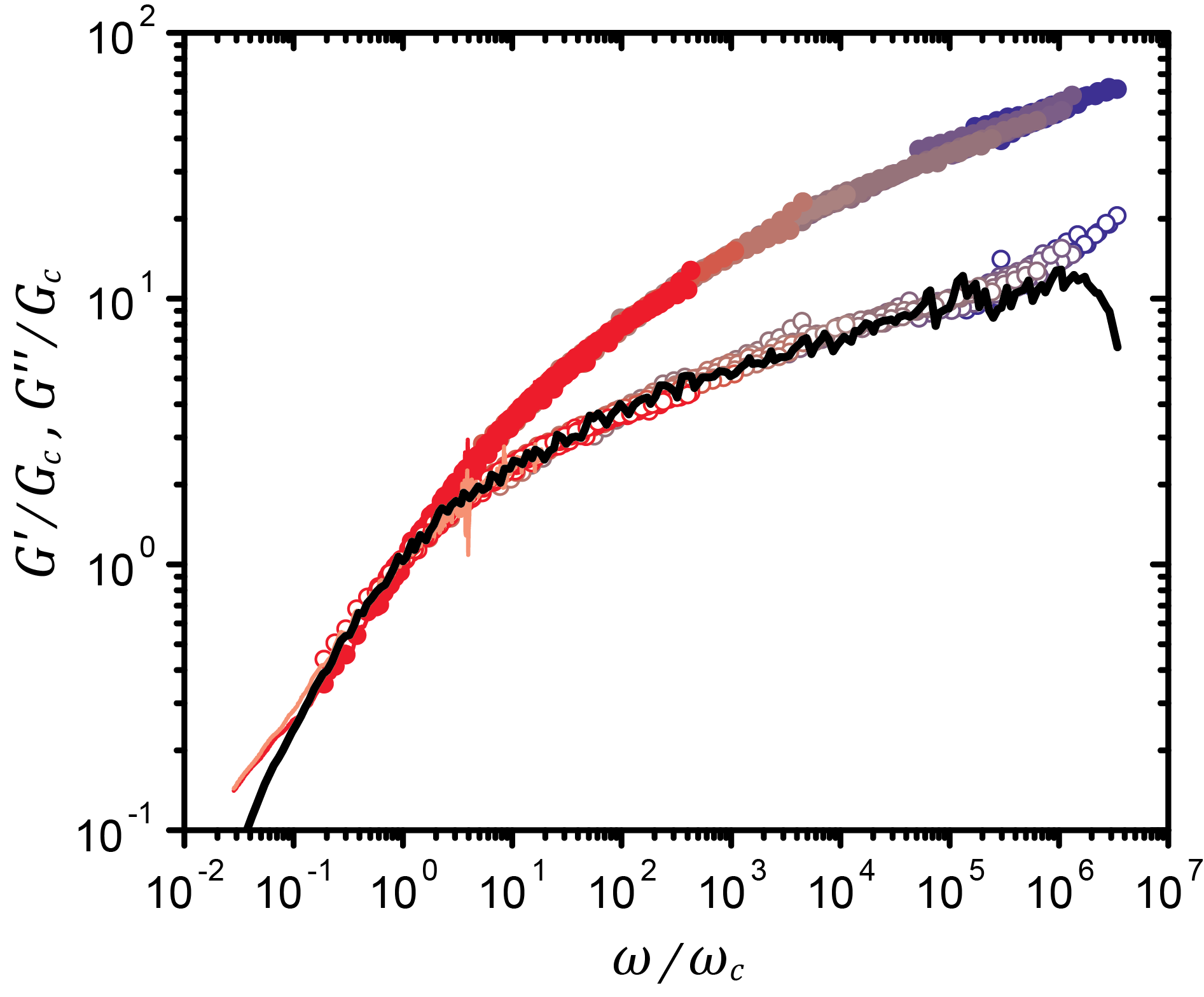}
\caption{\label{supp_mastercurve} Storage modulus $G'$ (filled) and loss modulus $G''$ (open) scaled with the crossover frequency $\omega_c$ in the abscissa and the crossover modulus $G_c$ in the ordinate for different mixing proportions $\chi$ at $\phi=5.0\%$ (same as Fig. 5 of the main text), and $G''/{G_c}$ calculated from $\abs{G^*(\omega)}/{G_c}=\sqrt{{G'(\omega)}^2+{G''(\omega)}^2}/{G_c}$ by the Kramers-Kronig relations (black line). The dark ($G'$) and light ($G''$) red lines at lowest frequencies represent the converted moduli from the creep compliance measured for $\chi=1$.}
\end{figure}

To verify the master curve of the scaled viscoelastic moduli $G'/{G_c}$ and $G''/{G_c}$, we ensure that the master curve conforms to the causality of linear responses by using the Kramers-Kronig relations \cite{Parot2007,Rouleau2013}. We first compute the scaled complex modulus $\abs{G^*(\omega)}/{G_c}=\sqrt{{G'(\omega)}^2+{G''(\omega)}^2}/{G_c}$, and convert it into the loss tangent $G''/G'$ using the Kramers-Kronig relations. Then we calculate $G''/G_c$ using the converted loss tangent and the experimentally obtained $G'/G_c$, and compare the result with the experimentally obtained $G''/G_c$. The two quantities exhibit good agreement, as shown in Fig.~\ref{supp_mastercurve}. The computed $G''/G_c$ at both ends deviates from the experimentally obtained master curve because of the finite range of the frequencies covered in our data. \par

\section{Sample micrographs}
\label{sample_img}
\begin{figure*}[t]
\setlength{\abovecaptionskip}{10pt}
\hspace*{-0.03cm}\includegraphics[scale=0.105]{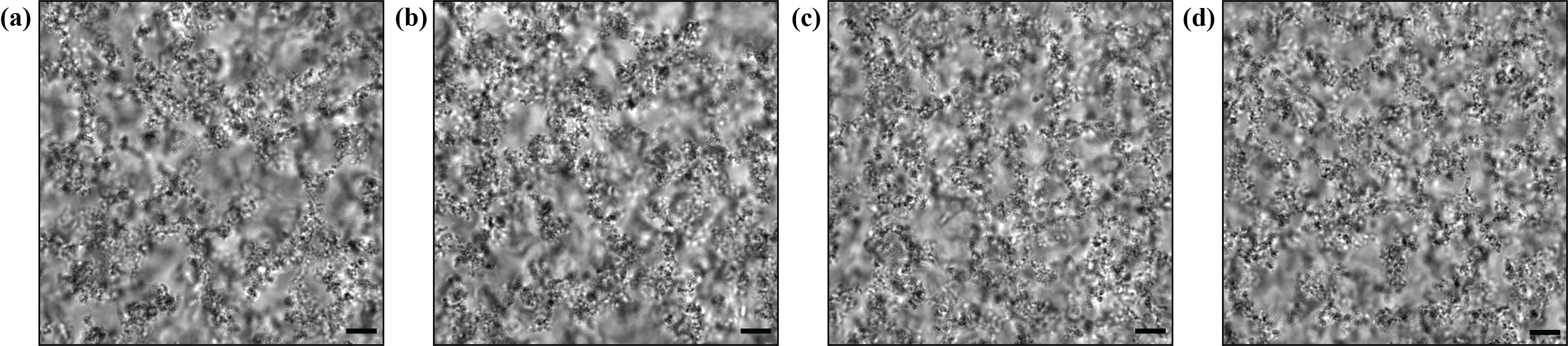}
\caption{\label{supp_gel_imag} Micrographs of the gels at $\phi=0.8\%$ for the mixing proportions $\chi=$ 0 (a), 0.5 (b), 0.7 (c), 1 (d). Scale bars in the bottom right-hand corners correspond to $10\;\si{\micro\meter}$, approximately equal to the cluster diameter.}
\end{figure*}

We display sample micrographs of the gels with increased contrast in Fig.~\ref{supp_gel_imag}. The fields of view of the images used for the calculation of the image structure function are four times as large as the ones in Fig.~\ref{supp_gel_imag}. Although evident in the static structure factor $S(q)$ in the reciprocal space, the fractal dimension $d_f$ is not clearly visible in the real-space micrographs, since the fractal structure appears over small length scales only. \par

\makeatletter\onecolumngrid@push\makeatother 
\clearpage
\makeatletter\onecolumngrid@pop\makeatother

\bibliographystyle{apsrev4-1}
\bibliography{Binary_gel_paper_Sept_20_mod.bib}

\end{document}